\documentclass[prb,twocolumn,floatfix,showpacs]{revtex4}
\usepackage{graphicx}
\begin{document}
\title{Level statistics of $XXZ$ spin chains with a random magnetic field}

\author{Kazue Kudo}
\affiliation{Graduate School of Humanities and Sciences, Ochanomizu
University, 2-1-1 Ohtsuka, Bunkyo-ku, Tokyo 112-8610, Japan}
\email{kudo@degway.phys.ocha.ac.jp}
\author{Tetsuo Deguchi}
\affiliation{Department of Physics, Ochanomizu University, 2-1-1
Ohtsuka, Bunkyo-ku, Tokyo 112-8610, Japan}
\email{deguchi@phys.ocha.ac.jp}

\date{\today}

\begin{abstract}
The level-spacing distribution of a spin-$\frac12$ $XXZ$ chain  
is numerically studied under random magnetic field.  
We show explicitly how the level statistics depends 
on the lattice size $L$, the anisotropy parameter $\Delta$, 
and the mean amplitude of the random magnetic field $h$. 
In the energy spectrum, quantum integrability competes with   
nonintegrability derived from the randomness,  
where the $XXZ$ interaction is modified by the parameter $\Delta$. 
When $\Delta \ne 0$, the level-spacing distribution mostly 
shows  Wigner-like behavior, while 
when $\Delta=0$, Poisson-like behavior appears  
although the system is nonintegrable due to randomness. 
Poisson-like behavior also appears for $\Delta \ne 0$
in the large $h$ limit.  
Furthermore, the level-spacing distribution  depends 
on the lattice size $L$, particularly
when the random field is weak.  
\end{abstract}

\pacs{75.10.Pq, 75.10.Jm, 05.30.-d, 75.10.Nr}

\maketitle

\section{Introduction}

Random matrix theories have been successfully applied to analysis of the
spectra of various physical systems such as quantum spin
systems,~\cite{Montam,Hsu,Poil,Pals,Angles,NNN}  
strongly correlated systems,~\cite{Faas}
and disordered quantum systems.~\cite{Shklov,Berkovits,Georgeot,Avishai} 
In quantum spin chains, if a given Hamiltonian is integrable by the Bethe
ansatz, the level-spacing distribution should be described by the
Poisson distribution:
\begin{equation}
 P_{\rm Poi}(s) = \exp (-s).
 \label{eq:Poisson}
\end{equation} 
If it is not integrable, the level-spacing distribution should be given
by the Wigner distribution: 
\begin{equation}
 P_{\rm Wig}(s) = \frac{\pi s}{2} \exp \left( - \frac{\pi s^2}{4}
                                           \right).
 \label{eq:Wigner}
\end{equation}
In the Anderson model of disordered systems, $P_{\rm Poi}(s)$ and
$P_{\rm Wig}(s)$ characterize the localized phase 
and the metallic phase, respectively.\cite{Shklov}

The numerical observations should be 
important.\cite{Montam,Hsu,Poil,Pals,Angles,NNN,Faas,Shklov,Berkovits,Georgeot,Avishai}    
In fact, there has been no direct theoretical derivation for the suggested 
behavior of the level-spacing distribution. Furthermore,  
unexpected behavior has been recently found 
in the level statistics of some $XXZ$ spin chains.~\cite{NNN} 
Robust non-Wigner behavior has been seen in the level-spacing distributions of
next-nearest-neighbor  coupled $XXZ$ chains,  
although they are nonintegrable. 
The reason why it appears is not clear yet, although 
we have considered two possible reasons: 
extra symmetries or finite-size effects.

In this paper, we discuss the level-spacing distribution 
of a disordered $XXZ$ spin chain so that we may find possible clues to 
the unexpected behavior of the $XXZ$ spin chains. 
We consider specifically the spin-${\frac 1 2}$ $XXZ$ spin chain 
with  random magnetic field, where quantum integrability 
competes with nonintegrability due to the randomness. 
The Hamiltonian on $L$ sites is given by 
\begin{equation}
 \mathcal{H} =J\sum_{j=1}^L \left( S^x_j S^x_{j+1} + S^y_j S^y_{j+1}  +
    \Delta S^z_j S^z _{j+1} \right) + \sum_{j=1}^L h_j S^z_j ,
\label{eq:H}
\end{equation}
where $S^{\alpha}=(1/2)\sigma^{\alpha}$ and
$(\sigma^x,\sigma^y,\sigma^z)$ are the Pauli matrices; $h_j$ is  random
magnetic field along the $z$ axis at site $j$; the periodic boundary
conditions are imposed. 
The random magnetic field $h_j$'s are
uncorrelated random numbers with a Gaussian distribution: 
$\langle h_j \rangle =0$ and 
$\langle h_n h_m \rangle  = h^2 \delta_{nm}$. 
Here we recall that the system (\ref{eq:H}) is integrable when $h=0$, 
while it is not when $h \ne 0$.

We show explicitly how the level-spacing distribution $P(s)$
depends on the lattice size $L$, the anisotropy parameter $\Delta$, 
and the mean amplitude of the random magnetic field $h$. 
For some special cases of the Hamiltonian (\ref{eq:H}), 
the level statistics has been discussed.  
The present numerical results should 
make explicit connections among them and even extend them.   
When $\Delta \ne 0$, $P(s)$ mostly shows Wigner-like behavior. 
For the Heisenberg case ($\Delta=1$), it has been shown 
that $P(s)$ coincides with $P_{\rm Wig}(s)$.~\cite{Avishai} 
When $\Delta=0$, Poisson-like behavior appears  
although the system is nonintegrable due to the randomness. 
However, the result is consistent with the Anderson localization of 
one-dimensional (1D) systems. The Wigner-like behavior for $\Delta \ne 0$
suggests that the Anderson localization in 1D systems 
should be broken by the interaction such as the 
$XXZ$ coupling. Here we note that an analogous phenomenon  
has been observed in the 2D Anderson model 
with electron interaction.~\cite{Berkovits}  
In the large $h$  and small $h$ limits, 
Poisson-like behavior appears again for $\Delta \ne 0$,  
which is consistent with the numerical results for 
spin-glass clusters,~\cite{Georgeot} 
the open-boundary Heisenberg chain,~\cite{Santos}
and the 3D Anderson model.~\cite{Shklov}
Furthermore, we find that $P(s)$ depends 
on the lattice size $L$, particularly
when the random field is weak.

There is another motivation for the present study. 
The symmetry of the integrable quantum $XXZ$ spin chain 
can  be nontrivial.  An extraordinary symmetry appears 
for special values of $\Delta$: 
the $XXZ$ Hamiltonian commutes with the $sl_2$ loop algebra 
when $q$ is a root of unity, where $q$
is defined  by $\Delta=(q+1/q)/2$. \cite{DFM}  
The loop algebra is an infinite-dimensional Lie algebra, 
and the dimensions of degenerate eigenspaces  
are given by some exponential functions 
of the system size, which can be extremely large. \cite{FM,Deguchi} 
It should therefore be nontrivial 
how the large degeneracies are resolved by nonintegrability.

\section{Numerical procedure}

In the Hamiltonian (\ref{eq:H}), total $S^z$ is conserved. The
eigenstates with different $S^z$ are uncorrelated. Therefore, we
consider only the largest subspace $S^z =0$. The largest sectors for
the lattice size $L=$8, 10, 12, 14 have 70, 252, 924, 3432 eigenvalues,
respectively (see Table \ref{tab:1}).

\begin{table}
 \caption{\label{tab:1}Matrix size of the largest subspace and the
 number of samples calculated in this work for each lattice size.}
\begin{ruledtabular}
 \begin{tabular}{ccr}
 Lattice size & Matrix size & Number of samples \\ 
\hline 
 8 & $70 \times 70$ & 10000 \\
10 & $252 \times 252$ & 3000 \\
12 & $924 \times 924$ & 1000 \\
14 & $3432 \times 3432$ & 900 \\
 \end{tabular}
\end{ruledtabular}
\end{table}

To find universal statistical properties of the Hamiltonians, one has to
deal with unfolded eigenvalues instead of raw eigenvalues. The
unfolded eigenvalues are renormalized values, whose local density of
states is
equal to unity everywhere in the spectrum. In this paper, the unfolded
eigenvalues $x_i$ are obtained from the raw eigenvalues $E_i$ in the
following method. Let us define the integrated density of states as
\begin{equation}
 n(E) = \sum _{i=1}^N \theta (E-E_i).
\end{equation}  
Here $\theta (E)$ is the step function and $N$ is the number of the
eigenvalues. We choose some points of coordinates: $(E_i, n(E_i))$ for
$i=1,21,41,\cdots,N$. The average of integrated density of states $\langle
n(E) \rangle$ is approximated by the spline interpolation through the
chosen points. The unfolded eigenvalues are defined as
\begin{equation}
 x_i = \langle n(E_i) \rangle .
\end{equation}
The level-spacing distributions are given by the probability function
$P(s)$, where $s=x_{i+1}-x_i$.

We have calculated 10000, 3000, 1000, 900 samples of $P(s)$ for $L=$8, 10,
12, 14, respectively (see Table \ref{tab:1}), and averaged the samples for
each $L$. To calculate the eigenvalues, 
we have used standard numerical
methods, which are contained in the LAPACK library.
 
\section{Level-spacing distribution}

\begin{figure}
\includegraphics[width=8cm]{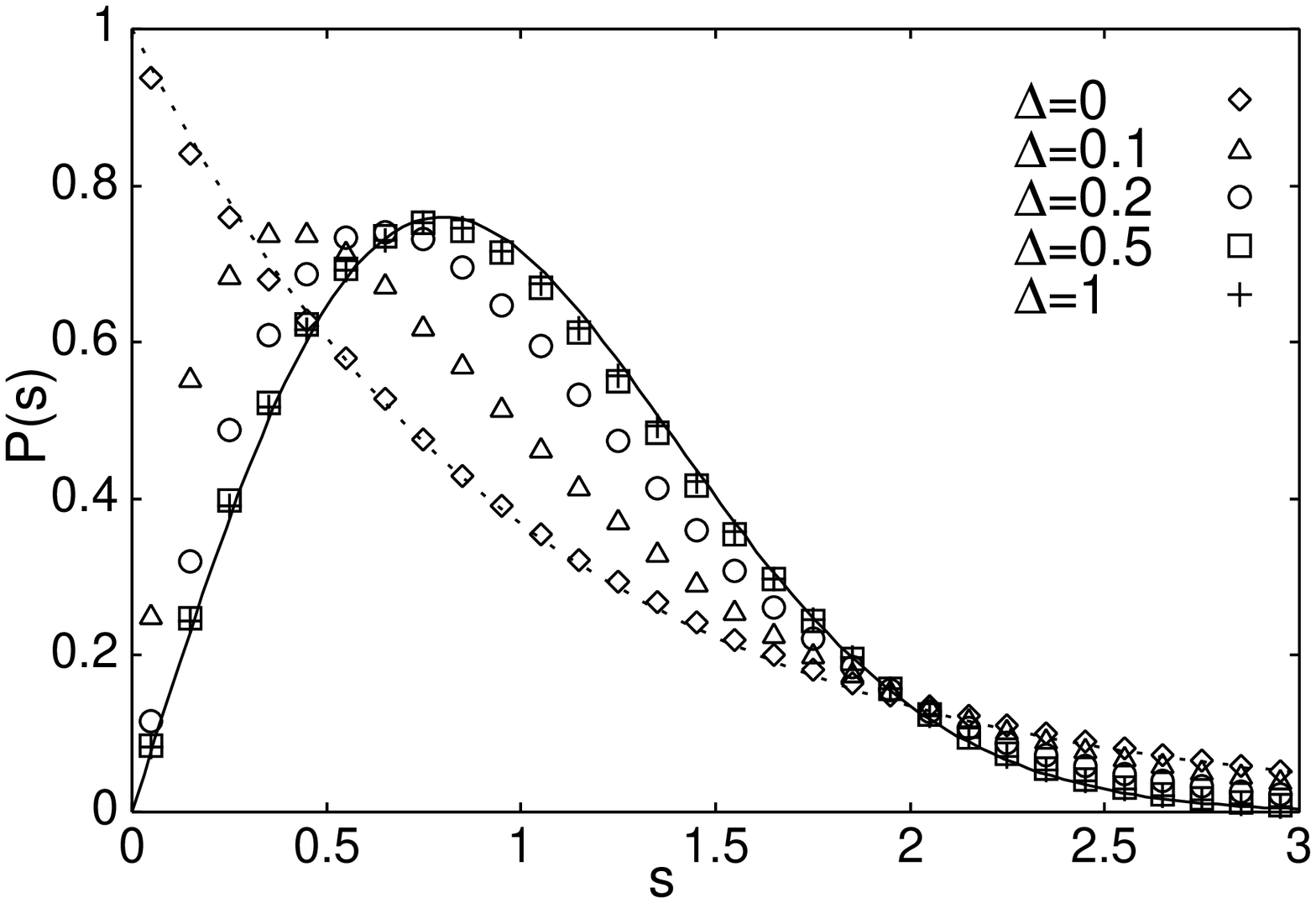}
\caption{\label{fig:1} Level-spacing distributions for $L=14$,
 $h/J=0.5$, $\Delta=$0, 0.1, 0.2, 0.5, 1. Broken line, the Poisson
 distribution; solid line, the Wigner distribution.}
\end{figure}

Depending on the anisotropic parameter 
$\Delta$ ($0 \le \Delta \le 1$), the level-spacing 
distribution $P(s)$ changes between the Wigner distribution 
$P_{\rm Wig}(s)$ and the Poisson distribution 
$P_{\rm Poi}(s)$ as shown in Fig.~\ref{fig:1}, where
$L=14$ and $h/J=0.5$. 
When $\Delta =0$, $P(s)$  almost coincides with 
$P_{\rm Poi}(s)$ although the system is nonintegrable 
due to the random magnetic field. As $\Delta$ increases, 
$P(s)$ rapidly changes to $P_{\rm Wig}(s)$.


Let us explain the Poisson-like behavior of $\Delta=0$ in terms of  
the Anderson localization.  
The Hamiltonian~(\ref{eq:H}) can be mapped into a model of interacting
1D free fermions under random potential:
\begin{eqnarray}
 \mathcal{H} &=& \frac{J}4 \cdot 2\left[ \sum^{L-1}_{j=1}
(c^{\dagger}_j c_{j+1}+ c^{\dagger}_{j+1} c_j)
  -(-1)^M (c^{\dagger}_L c_1 + c^{\dagger}_1 c_L) \right] \nonumber\\
&+& \Delta  \frac{J}4 \sum^L_{j=1}
(4c^{\dagger}_j c_j c^{\dagger}_{j+1} 
c_{j+1} -4c^{\dagger}_j c_j +1)\nonumber\\
&+& \sum^L_{j=1} h_j\left(\frac12 - c^{\dagger}_j c_j \right).
\label{eq:JW}
\end{eqnarray} 
Here, $L$ is the number of sites; $M$ is the number of fermions; 
$c^{\dagger}_j$ and $c_j$ are the creation and annihilation
operators of fermions on the $j$th site, respectively. And the Anderson
model of noninteracting disordered fermions is given by
\begin{equation}
 \mathcal{H}=\sum_j \varepsilon_j c^{\dagger}_j c_j
 +\sum_{\langle i,j \rangle} V (c^{\dagger}_i c_j + c^{\dagger}_j c_i),
\label{eq:Anderson}
\end{equation}
where $\varepsilon_j$ is the random potential at the $j$th site; $V$ is
a constant hopping integral; $\langle i,j \rangle$ denotes summation
over nearest-neighbor sites. One can find that Eq.~(\ref{eq:JW}) for
$\Delta=0$ corresponds to Eq.(\ref{eq:Anderson}) of the 1D case.
It is known that localization always occurs in the 1D case, while
the 3D Anderson model has the metallic 
phase and the localized phase.  
Here we recall that the metallic phase corresponds to $P_{\rm Wig}(s)$ 
and the localized phase to $P_{\rm Poi}(s)$. 
Thus, the observed Poisson-like behavior for $\Delta=0$ is consistent with 
the Anderson localization.

Let us discuss a consequence of the Wigner-like behavior for 
$\Delta \ne 0$. 
The Hamiltonian~(\ref{eq:H}), namely Eq.~(\ref{eq:JW}), 
for $\Delta \ne 0$ corresponds to 
interacting 1D fermions under random potential. 
Thus, 
the Wigner-like behavior of $P(s)$ for $\Delta \ne 0$ might suggest that 
the interaction among fermions 
should break the Anderson localization in 1D chains. 
The suggestion could be consistent with the observation 
on the 2D interacting lattice fermions where 
the level-spacing distribution 
changes from $P_{\rm Poi}(s)$ 
to $P_{\rm Wig}(s)$ as the electron-electron  
interaction increases from zero.~\cite{Berkovits} 

\begin{figure*}
\includegraphics[width=16cm]{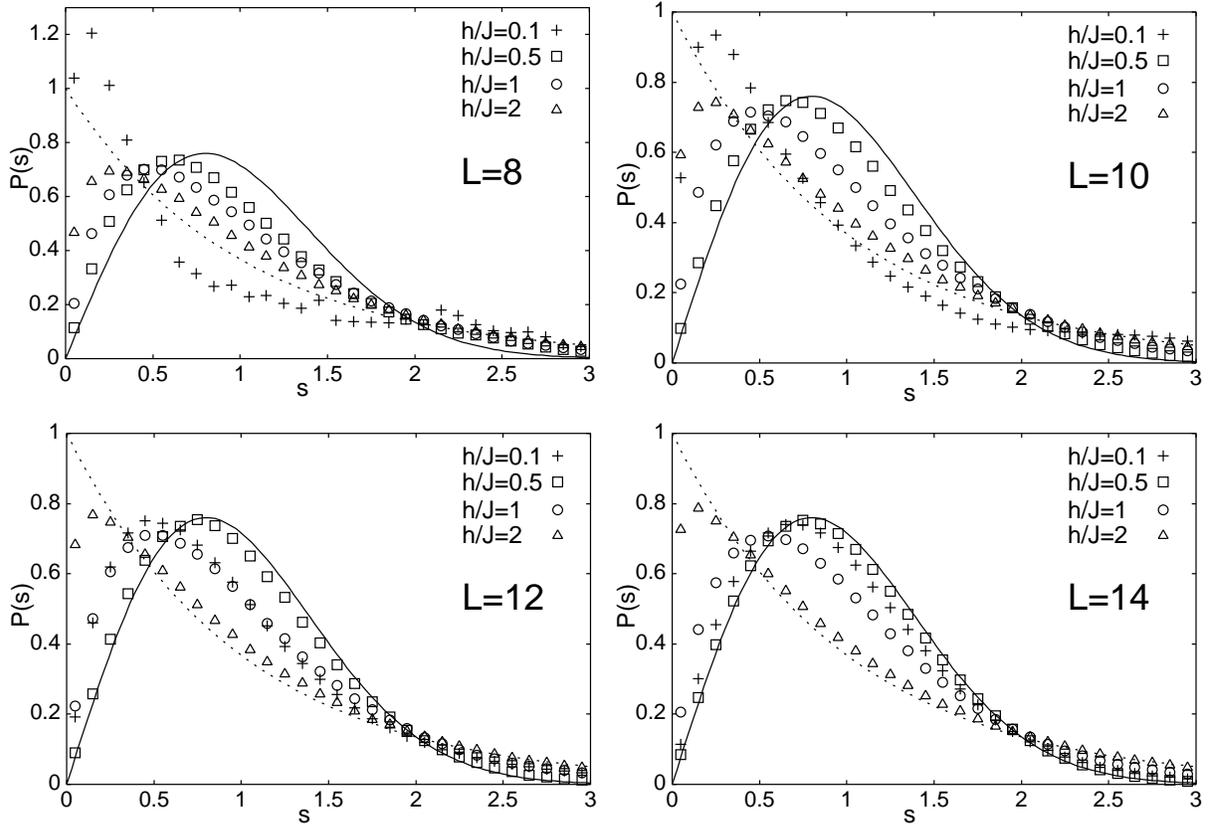}
\caption{\label{fig:2} Level-spacing distributions for $L=$ 8, 10, 12,
 14, $h/J=$0.1, 0.5, 1, 2, $\Delta=0.5$. 
 Broken lines, the Poisson distribution; solid lines, the Wigner
 distribution. }
\end{figure*}

We now discuss how the level-spacing distribution $P(s)$ 
depends on the random magnetic field $h$. 
We consider only the case of $\Delta \ne 0$.  
In Fig.~\ref{fig:2} the graphs of $P(s)$ are shown 
for some values of $h/J$ and $L$, where $\Delta=0.5$. 
We first consider the case of large $h$. 
As $h/J$ increases from the value of 0.5,  
we observe that the form of $P(s)$ changes from 
$P_{\rm Wig}(s)$ to $P_{\rm Poi}(s)$. The observation suggests that  
the effect of random magnetic field on each site should become
larger than that of the correlation between adjacent spins, 
as the random field $h/J$ increases. 
The spins should become more independent of each other as $h/J$ increases, 
since the effect of correlation decreases effectively. 
Thus, the Poisson-like behavior 
of $P(s)$ should appear in the limit of  
large $h/J$.  Similar shifts 
from $P_{\rm Wig}(s)$ to $P_{\rm Poi}(s)$ 
as randomness increases 
have been discussed for the 3D Anderson model,~\cite{Shklov} 
the spin-glass clusters,~\cite{Georgeot} and the  
open-boundary Heisenberg chain.~\cite{Santos}

For the case of small $h$, the level-spacing distribution $P(s)$
strongly depends on the system size $L$, 
and the behavior of $P(s)$ is dominated by finite-size effects.
In Fig. 2,  we observe that the form of 
$P(s)$ for $h/J=0.1$ is different from that of 
the standard Wigner distribution, particularly 
when $L$ is small.
When $L$ is small, random magnetic field is irrelevant to energy levels
if it is smaller than the order of $1/L$. In fact, energy differences
should be at least in the order of $1/L$, and random magnetic field can be
neglected if it is much smaller than some multiple of $1/L$.
Thus, for
the case of small $h$, 
the level statistics should show such a behavior as that of $h=0$. 
In fact, the Hamiltonian for $h=0$ is the
integrable $XXZ$ spin chain, which should have Poisson-like behavior. 
Furthermore, the integrable $XXZ$ Hamiltonian at $\Delta=0.5$  
has the $sl_2$ loop algebra symmetry,~\cite{DFM}   
and the level-spacing distribution should show a peak at $s=0$.~\cite{NNN} 
In Fig. 2, the graph of $P(s)$ for $L=8$ and $h/J=0.1$ 
suggests such behavior.

\begin{figure}
\includegraphics[width=8cm]{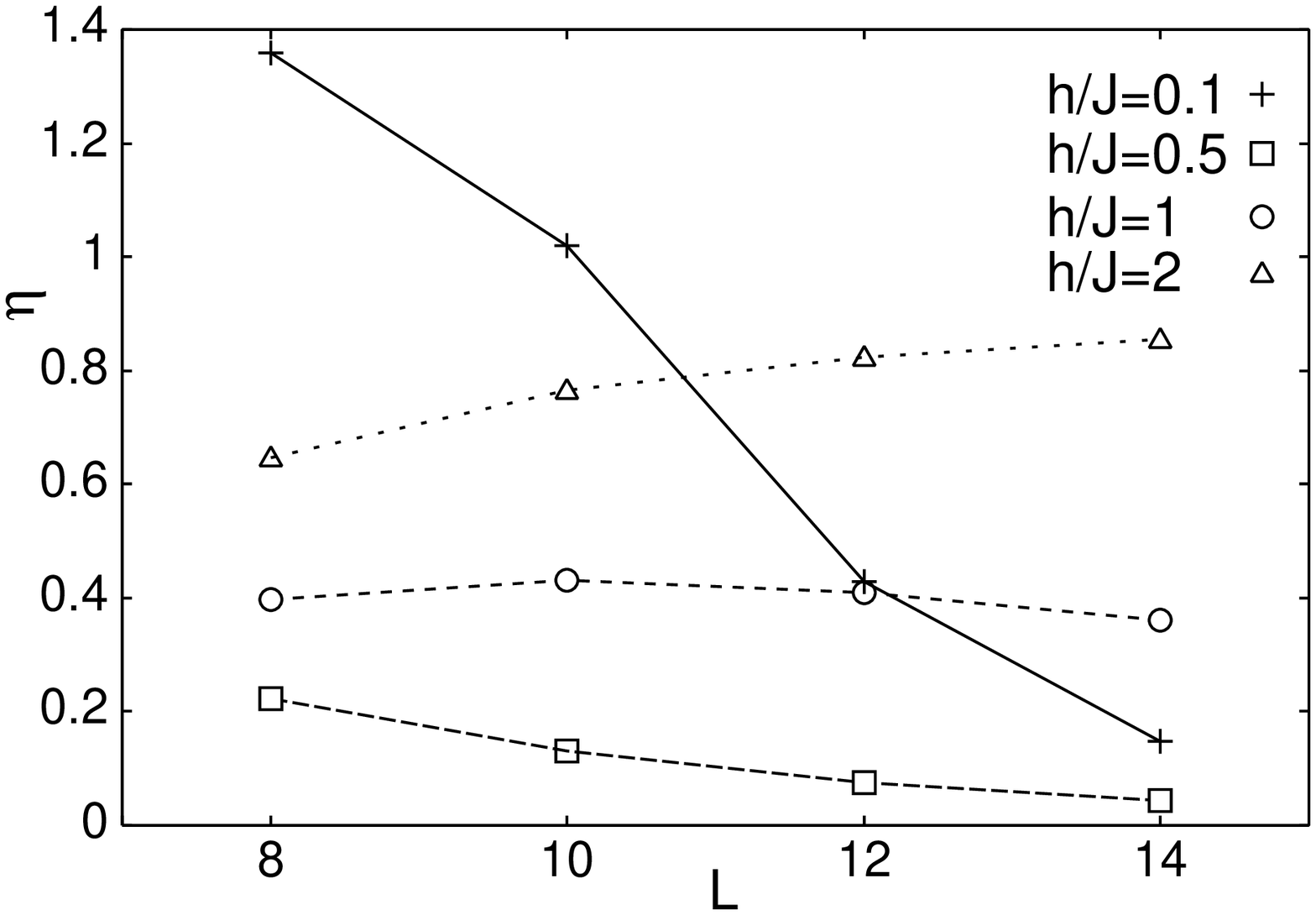}
\caption{\label{fig:3} Dependence of the parameter $\eta$ on the
 lattice size $L$ for $h/J=$0.1, 0.5, 1, 2   
and $\Delta=0.5$. $\eta=0$ corresponds to the Wigner
 distribution and $\eta=1$ to the Poisson distribution.}
\end{figure}

Let us discuss finite-size effects 
on the level-spacing distributions. 
In order to observe the size dependence of $P(s)$  clearly, 
we employ the following parameter,  
$\eta = \int_0^{s_0} [P(s)-P_{\rm Wig}(s)] ds /
\int_0^{s_0} [P_{\rm Poi}(s)-P_{\rm Wig}(s)] ds $, where 
$s_0=0.4729 \cdots$ is the intersection point of $P_{\rm Poi}(s)$ and 
$P_{\rm Wig}(s)$.~\cite{Georgeot,Santos} 
Thus, we have $\eta=0$ when $P(s)$ coincides with 
$P_{\rm Wig}(s)$, and $\eta=1$ when $P(s)$ coincides with 
$P_{\rm Poi}(s)$. 
In Fig. 3,  the value of $\eta$ for $h/J=0.1$ 
strongly depends on the lattice size $L$. Moreover, we observe 
that as $L$ increases, $\eta$ decreases for $h/J=0.5$,  
while $\eta$ increases for $h/J=2$. 
The observation suggests that 
$\eta$ approaches either the value 0 or 1 as $L$ increases. 
In other words, it should become more definite  
whether $P(s)$ has Wigner-like behavior or not, 
as the system size becomes large.

\section{Conclusions}
In conclusion, we have calculated the level-spacing distributions of
finite spin-$\frac12$ $XXZ$ chains under random magnetic field, and
shown how the level-spacing distributions change between the Poisson
distribution and the Wigner distribution depending on the lattice size $L$,
the anisotropy parameter $\Delta$, and the mean amplitude of the random
magnetic
field $h$. For $\Delta=0$, the level-spacing
distribution $P(s)$ almost coincides with the Poisson distribution 
although the system has the randomness. 
As $\Delta$ increases from zero, 
$P(s)$ rapidly shifts to
the Wigner distribution.
The behaviors of $P(s)$ have been explained in terms of Anderson 
localization. 
For $\Delta \ne 0$, $P(s)$ strongly depends on $L$ when $h$ is small.
When $L$ is finite, $P(s)$ should show Poisson-like
behavior in the small $h$ limit.
In the large $h$ limit, however, $P(s)$ should become close to
the Poisson distribution independent of $L$. 

\begin{acknowledgments}
The authors would like to thank K.~Nakamura and T.~Kato for useful
 discussions. 
 The present study was partially supported by the Grant-in-Aid for
 Encouragement of Young Scientists (A):Grant No.~14702012.
\end{acknowledgments}

\end{document}